\documentclass[
superscriptaddress,
 amsmath,
 amssymb,
 aps,
 pra,
 twocolumn,
floatfix,
]{revtex4-2}

\usepackage{graphicx}
\usepackage{dcolumn}
\usepackage{bm}
\usepackage{physics}
\usepackage{natbib}

\begin{document}


\title{Loss compensated and enhanced mid-infrared\\interaction-free sensing with undetected photons}

\author{Nathan~R.~Gemmell}
\email{n.gemmell20@imperial.ac.uk}
\author{Jefferson~Fl\'orez}%
\author{Emma~Pearce}%
\author{Olaf~Czerwinski}%
\author{Chris~C.~Phillips}%
\author{Rupert~F.~Oulton}%
\affiliation{%
 Department of Physics, Blackett Laboratory, Imperial College London, South Kensington Campus, London SW7 2AZ, United Kingdom
}%

\author{Alex~S.~Clark}
\affiliation{%
 Department of Physics, Blackett Laboratory, Imperial College London, South Kensington Campus, London SW7 2AZ, United Kingdom
}%
\affiliation{
 Quantum Engineering Technology Labs, H. H. Wills Physics Laboratory and Department of Electrical
and Electronic Engineering, University of Bristol, BS8 1FD, United Kingdom
}%

\date{\today}

\begin{abstract}
Sensing with undetected photons enables the measurement of absorption and phase shifts at wavelengths different from those detected. Here, we experimentally map the balance and loss parameter space in a non-degenerate nonlinear interferometer, showing the recovery of sensitivity despite internal losses at the detection wavelength. We further explore an interaction-free operation mode with a detector-to-sample incident optical power ratio of >200. This allows changes in attowatt levels of power at 3.4~$\mu$m wavelength to be detected at 1550\,nm, immune to the level of thermal black-body background. This reveals an ultra-sensitive infrared imaging methodology capable of probing samples effectively `in the dark'.
\end{abstract}

\maketitle


\section{\label{sec:level1}Introduction}

Induced coherence without induced emission was first demonstrated experimentally by Zou, Wang, and Mandel in 1991\cite{Zou1991}, where interference was observed between the down-converted photons generated in two separate nonlinear crystals. It was shown that the interference was dependent on the mutual indistinguishability of the two crystals’ signal and idler fields; any distinguishability introduced between the idler fields, for example, reduced the visibility of interference between the signal beams at the detector. Remarkably, this allows information about an object probed by one idler beam to be gained only from measuring the interference between the signal beams. Whilst the first practical demonstration of an application of induced coherence was for spectroscopy\cite{Kulik2004}, its application to imaging has since caused a significant pique in interest\cite{Lemos2014}. This technique does not rely on up-conversion\cite{Barh2019, Kehlet2015} (which is often inefficient and can introduce spurious background noise) or indeed any detection capability at the probe wavelength, as is necessary for non-degenerate ‘ghost imaging’ \cite{Aspden2015, Chan2009}. The primary attraction to the imaging community is clear: infrared imaging capability using significantly cheaper, lower noise, and more readily available visible photon detection technology.  This capability has dramatic consequences in the field of mid-IR imaging where not only is imaging technology limited, but the thermal black-body background plays a significant role in the achievable signal-to-noise ratio. With induced coherence-based mid-IR imaging, background thermal noise only contributes at the signal wavelength, where it can be considered irrelevant for room temperature samples.

\begin{figure*}[t]
\includegraphics[width=13cm]{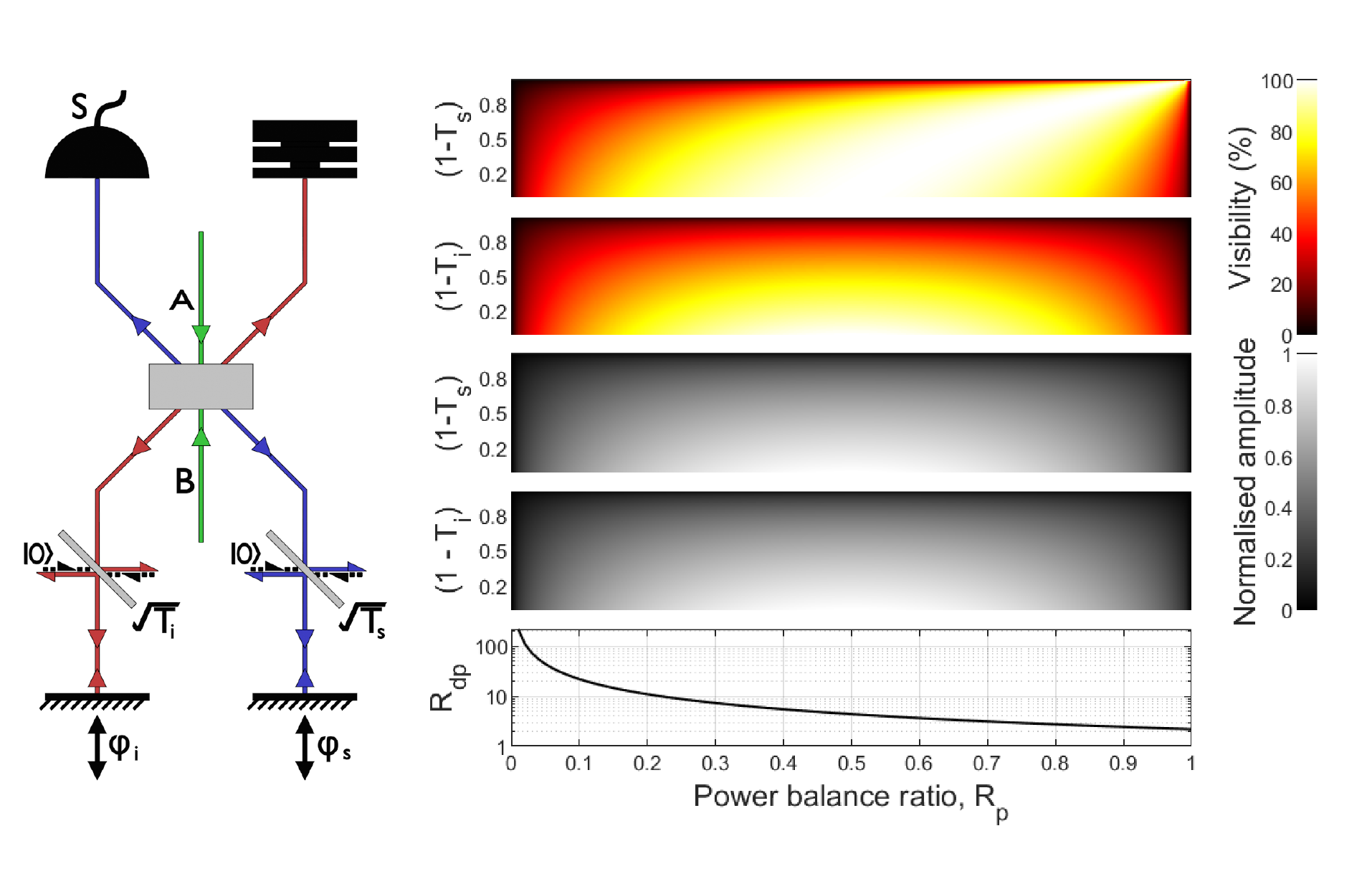}
\caption{Left: Schematic of theoretical model of a non-degenerate nonlinear interferometer. Green indicates pump beam, blue and red represent signal and idler beams, respectively. Process A is unseeded. Detection takes place on the signal beam after reflection and return through the nonlinear process overlapping with pump B and reflected idler beam. The output idler photons are discarded. Losses are modelled as beamsplitters with transmittivity of $\sqrt{T_\text{s/i}}$, such that total transmittivity of the beam is $T_\text{s/i}$. Phase shifts ($\phi_\text{s/i}$) are introduced on reflection. Right: Visibility ($\mathcal{V}$) and amplitude ($\mathcal{A}$) (normalised to input power) of the interference as a function of both internal signal/idler loss $(1-T_\text{s/i})$ on y-axis and power balance ratio $R_\text{p}$ on the x-axis. Bottom plot shows the detection-to-probe power ratio $R_\text{dp}$.}
\label{fig:theory}
\end{figure*}

Much effort in recent years in the field of induced coherence has involved simplifying the methodology and apparatus to make the technique more accessible. This has resulted in the implementation of `folded' systems which use a double pass of a single nonlinear crystal\cite{Basset2021, Kviatkovsky2020, Paterova2018, Lindner2020}, a seeded version of the experiment \cite{Cardoso2018}, an interferometer that uses two different nonlinear processes\cite{Vergyris2020}, and spectroscopy based on all fields passing stacked crystals \cite{Kalashnikov2016}. This work continues to push the technology closer to real-world applications \cite{Buzas2020}. However, practical imaging systems will require tunability and flexibility in their sensitivity and dynamic range if they are to make a real impact. For example,  the so-called ‘high-gain’ regime (where pump powers are high enough for the down-conversion process to self-seed) offers a pathway to a tunable dynamic range\cite{Kolobov2017} by matching the interferometer’s response range to a sample’s transmissivity, in addition to offering greater phase sensitivity \cite{Giese2017, Kolobov2017}. Not discounting the limitations of pump depletion\cite{Florez2020}, this tunability currently requires operation with unrealistic levels of power for some practical imaging systems. Nonetheless, the balance of photon numbers generated in the two processes of a nonlinear interferometer is central to maximising its effective dynamic range.

Imaging systems based on induced coherence are also ideal for photosensitive samples as the object need only be interrogated by the idler beam from the first process while the detector sees the induced coherence via signal beam photons from two. This leads to an effect akin to `interaction-free imaging'\cite{White1998, Elitzur1993}, where there can be a discrepancy between the optical power impinging on the sample and that which hits the detector (since the detected photons do not interact with the sample). In the case of a non-degenerate nonlinear interferometer, this effect is enhanced due to the mismatch in the signal and idler wavelengths (where the beam that probes a sample has a significantly lower photon energy and is therefore inherently less photo-damaging). This power discrepancy seen between sample and detector can be enhanced even further by unbalancing the interferometer, such that even fewer photons are generated in the path that interacts with the sample. Again, careful balancing of the interferometer is important to maintain detectable interference visibility.

In this work, we study the balance of power in nonlinear interferometers both theoretically and experimentally. We show that losses of the first signal beam within the interferometer can be compensated for and that the `interaction-free' imaging effect can be exploited. This could have significant implications for the design of future nonlinear interferometers, particularly those based on ‘folded’ designs where redressing the balance of the system is more difficult, or those designed for probing ultra-photosensitive samples.

\begin{figure*}[t]
\includegraphics[width=16cm]{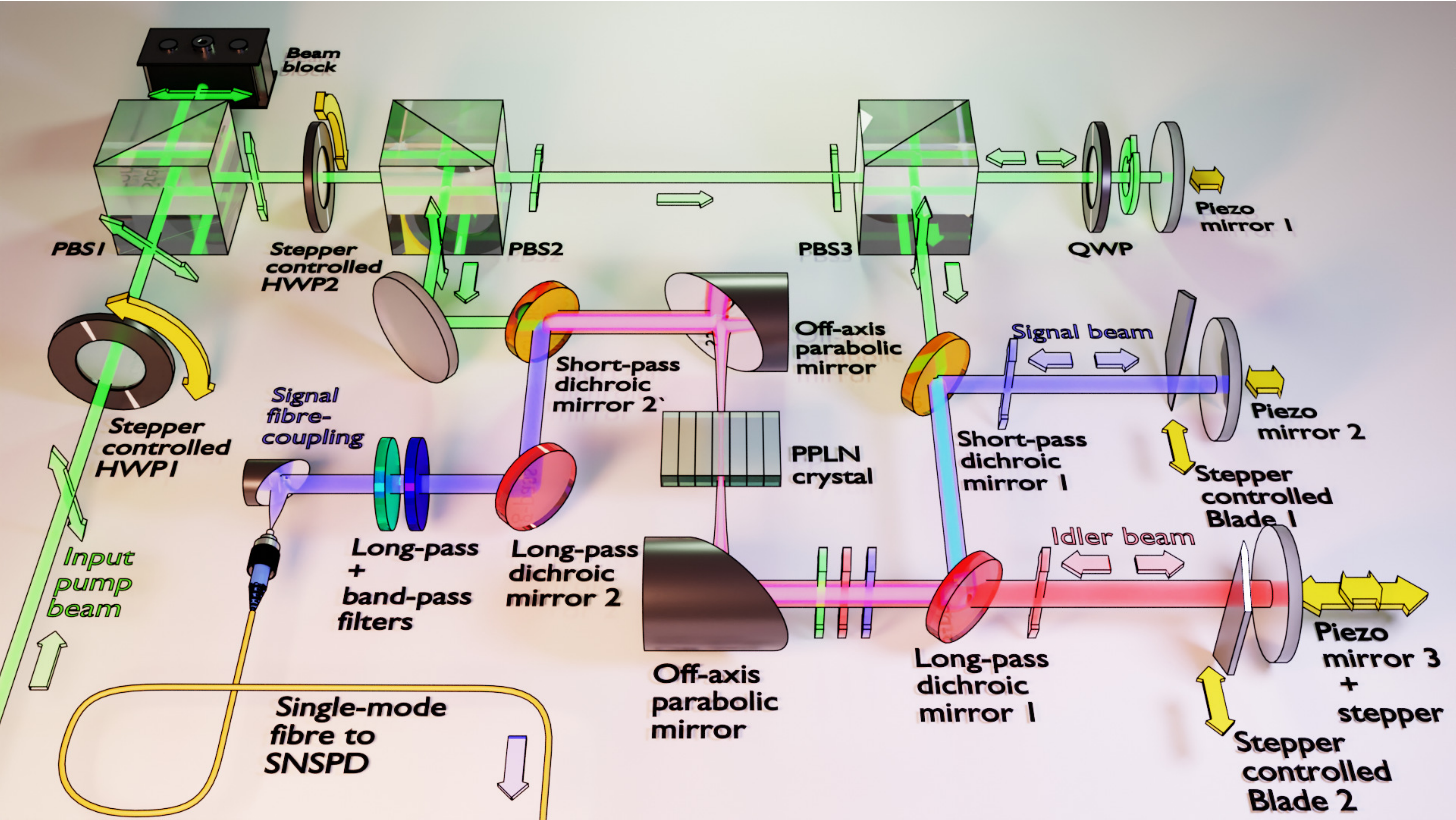}
\caption{Schematic of experimental setup. Green beam paths indicate pump (1064\,nm), blue represents the signal paths (1550\,nm), and red corresponds to the idler (3.4\,$\mu$m). Yellow arrows indicate automated translation.}
\label{fig:setup}
\end{figure*}

\section{\label{sec:level2}Theory}
We restrict our theoretical analysis only to the Michelson 'folded' design. In such an interferometer, a single nonlinear crystal is pumped in one direction, generating the first signal and idler fields at wavelengths $\lambda_\text{s}$ and $\lambda_\text{i}$. These are subsequently reflected back through the crystal with a second pump (most commonly the first pump reflected), generating the second signal and idler fields necessary for the induced coherence. Such a system can be comprehensively modelled using the theory presented by E.~Giese et al. \cite{Giese2017}, and it is this framework we shall be using to analyse our system. We also restrict any analysis to the non-degenerate case since this is where the majority of interest lies for practical imaging applications, making use of the transfer of information between wavelengths. Further, since detector losses merely act as a multiplying term, we omit them here. A schematic of the model is shown in Fig.\,\ref{fig:theory} where A and B in this case denote the first and second passes of the same nonlinear crystal. From this treatment, it can be shown that the number of photons incident on our signal detector can be expressed as
\begin{multline}
    N_\phi = T_\text{s}G_\text{A} + T_\text{i}G_\text{B} + (T_\text{s} + T_\text{i})G_\text{A}G_\text{B}\\
    - 2\sqrt{T_\text{s}T_\text{i}(1+G_\text{A})(1+G_\text{B})G_\text{A}G_\text{B}}\cos\phi \, ,
\label{eq:1}    
\end{multline}
where $T_\text{s}$ and $T_\text{i}$ are the transmissivities of the paths of the first signal and idler fields to the second crystal, respectively, and $G_\text{A}$ and $G_\text{B}$ represent hyperbolic functions defining the nonlinear gain parameters for the nonlinear crystals A and B (proportional to the input power via $G_\text{A/B}\propto\sinh^{2}(\sqrt P_\text{A/B})$. The parameter $\phi=\phi_\text{p} - \phi_\text{i} - \phi_\text{s}$ represents the total accumulated phase shift between the secondary pump and the primary signal and idler fields, respectively. Using Eq.\,\eqref{eq:1}, we can now model a nonlinear interferometer for arbitrary losses in the signal and idler beams, as well as for uneven gain parameters in the two crystals, due to unbalanced pumping powers.
The performance of nonlinear interferometers can be defined using two parameters: the amplitude and visibility of the interference fringes. The amplitude can be simply defined as the difference between the numbers of photons detected at the peak of the fringe (at $\phi_\text{max}$) and the trough (at $\phi_\text{max}+\pi$) such that $\mathcal{A} = N_{\phi_\text{max}} - N_{\phi_\text{max} + \pi}$. The visibility of the interference fringes, $\mathcal{V}$, is defined as the amplitude normalised by the sum of photon numbers at the peak and trough
\begin{equation}
    \mathcal{V} = \frac{\mathcal{A}}{\left(N_{\phi_\text{max}} + N_{\phi_\text{max} + \pi}\right)}  \, .
\end{equation}
In assessing the practicality of these systems as mentioned above, one can also measure the `interaction-free' efficiency simply defined as the number of photons incident on the detector over the number of photons incident on the sample. However, in a non-degenerate regime---such as that discussed here---there is also the energy discrepancy between the signal and idler photons leading to a further power unbalancing between the detected and probe light beams. This can then be defined as the detection-to-probe power ratio
\begin{equation}
    R_\text{dp} = \frac{N_\text{d}\lambda_\text{i}}{N_\text{s}\lambda_\text{s}} \, ,
\end{equation}
\noindent where $N_\text{d}$ and $N_\text{s}$ are the average number of photons incident on the detector and sample, respectively. The effects of power balance (ratio of input power to first and second nonlinear processes) and first signal beam transmission, $T_\text{s}$, on visibility and amplitude, are shown on the right in Fig.\,\ref{fig:theory}, along with a plot showing the detection-to-probe power ratio. The power balance $R_\text{p}$ used on the x-axis of these maps assumes a constant input power into the interferometer, which can be distributed arbitrarily between the two passes of the crystal and is simply calculated as the ratio of input power sent to the first pass to the total input power. All remaining power is assumed to be used in the second pass with no loss.

\begin{figure*}[t]
\includegraphics[width=14cm]{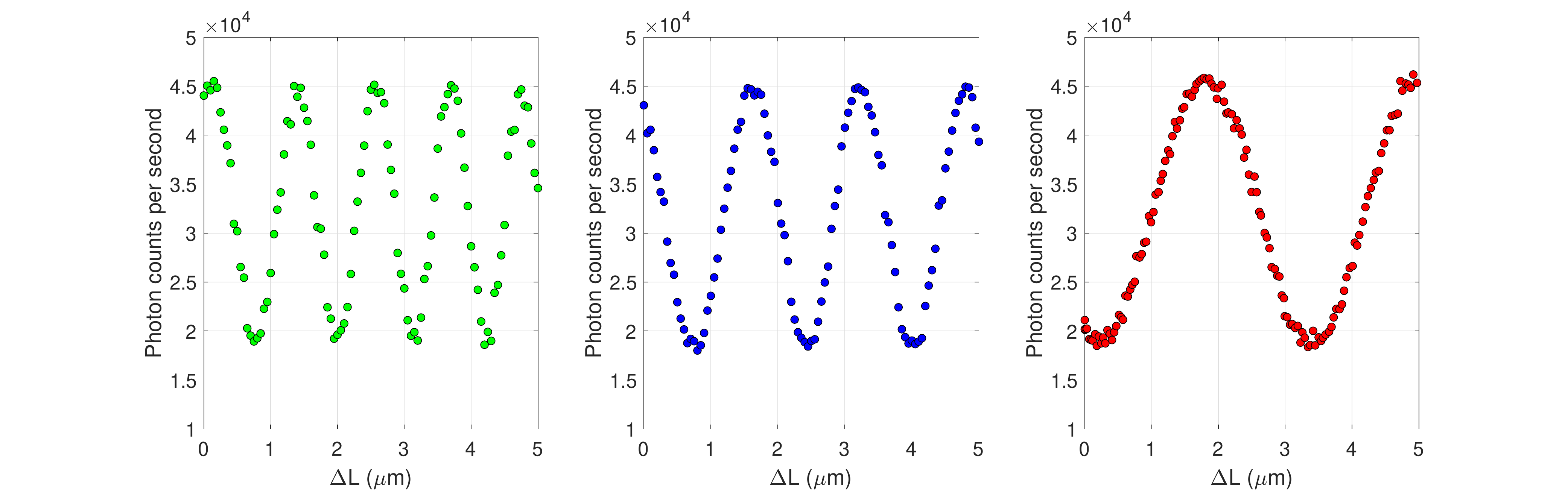}
\caption{Signal count rates as a function of the length change $\Delta$L in the secondary pump path (left, green), internal signal path (middle, blue), and internal idler path (right, red).}
\label{fig:fringes}
\end{figure*}

The top map shows the possibility to regain high visibility, even in the presence of high loss of the first signal beam, by simply directing more power to the first nonlinear process as shown by the asymmetry. This can be intuitively interpreted by considering that an equal probability of detecting a photon from either arm yields maximum visibility. However, as shown in the corresponding amplitude plot, sending more of the available power to the first process to maximise visibility comes at the expense of the amplitude of the interference fringes, as it will significantly reduce the amount of detectable signal. Unbalancing in the opposite direction (more power to the second process) dramatically increases the detection-to-probe power ratio, though this will always come at the expense of both amplitude and visibility. Losses in the first idler always result in a reduction of both signal and visibility and cannot be redressed with an unbalancing of pumps (within the low gain regime tested experimentally here), since this loss in the idler does not represent a change in the balance of signal photons generated in the first and second nonlinear processes.

\section{\label{sec:level3}Experimental procedure}
Figure\,\ref{fig:setup} shows a schematic of the unbalanced nonlinear interferometer designed to test this theory. The input power of the pump beam (1064\,nm Azurlight continuous-wave (CW) attenuated to $\sim1$\,mW) is adjusted via reflection from a polarizing beam splitter (PBS 1) after passing a half-wave plate (whose angle is computer-controlled by a servo motor, HWP 1). 
The pump beam is then split into two arms via a second PBS (PBS 2), that defines the two pump beams for the interferometer. The balance of power between the two pump processes can then be adjusted via a second servo-controlled half-wave plate (HWP 2), without changing the total power input to the system. This also allows an experimental study of the regime in which the parametric gain of the second process is higher than the first, something only previously studied with a two-crystal setup and a difference in phase-matching conditions\cite{Manceau2017}.

The second pump arm is sent through another PBS  (PBS 3) and a quarter-wave plate (QWP) before being reflected from a mirror to double pass the QWP, rotating the polarization to vertical and exiting the reflection port of the PBS. This reflection scheme enables the second pump phase to be adjusted by a computer-controlled piezo-electric motor (piezo mirror 1).

The nonlinear crystal is a temperature-controlled periodically-poled lithium niobate (PPLN) chip supplied by Covesion, with a poling period of $\Lambda$ = 30.5\,$\mu$m, and anti-reflection coated to R $<$ 1.5\% at 1064\,nm, to R $<$ 1\% between 1400 and 1800\,nm, and to R $\sim$6\%-3\% between 2600-4800\,nm, on both input/output facets. Both pump beams are weakly focused into either facet of the crystal through 152\,mm focal length silver off-axis parabolic mirrors (resulting in a 0.12\,mm $1/e^{2}$ pump beam waist). These mirrors also serve to collect and collimate the generated signal and idler photons from both pumping directions. Signal and idlers from the primary pump pass are tapped off from the pump beam via two dichroic mirrors (1180\,nm short-pass from Thorlabs (short-pass dichroic mirror 1)  and 2100\,nm long-pass from Layertec (long-pass dichroic mirror 1), respectively), and sent to two mirrors with computer-controlled piezo-motor positioning which reflect the beams back through the crystal with an accumulated phase shift from the piezo-motor (piezo mirror 2 \& 3). The idler mirror has an additional computer-controlled translation stage (stepper motor) to enable the path length between signal and idler photons to be matched. Two independent computer-controlled translation stages (stepper motors) adjust the positions of knife blades in the first signal and idler beams, allowing arbitrary amounts of loss to be introduced into each arm.
Spontaneous parametric down-conversion photons generated from the first pump returning through the crystal overlap with those generated from the second pump beam, with the signal photons subsequently being isolated via a second 1180\,nm shortpass dichroic (short-pass dichroic mirror 2). The signal photons are then filtered (2100\,nm long-pass from Layertec (long-pass dichroic mirror 2), and 12\,nm wide 1550\,nm bandpass and 1400\,nm longpass, from Thorlabs), coupled into standard telecommunications single-mode fibre (SMF-28) and sent to a superconducting nanowire single-photon detector (SNSPD, IDQuantique, $\sim80$\% detection efficiency at 1550\,nm), where count rates are monitored by a pulse counting module (ID900) with a 50\,ms acquisition time.

\section{\label{sec:level4}Results}
Figure\,\ref{fig:fringes} shows the fluctuation of single-photon signal beam counts as a function of the phase shift introduced by the piezo mirrors in the second pump, the signal, and the idler beams. As each piezo motor is scanned, the signal count rate oscillates with a period of the wavelength of the scanned field (1064\,nm for the pump, 1550\,nm for the signal and 3390\,nm for the idler). The idler piezo motor has a strain gauge (unlike the signal and pump piezos) which ensures that errors in the piezo position due to hysteresis can be monitored and corrected, and thus the idler wavelength can be measured to higher precision. For the rest of the results presented in this paper, it was the idler field that was scanned.

\begin{figure}
\includegraphics[width=8.3cm]{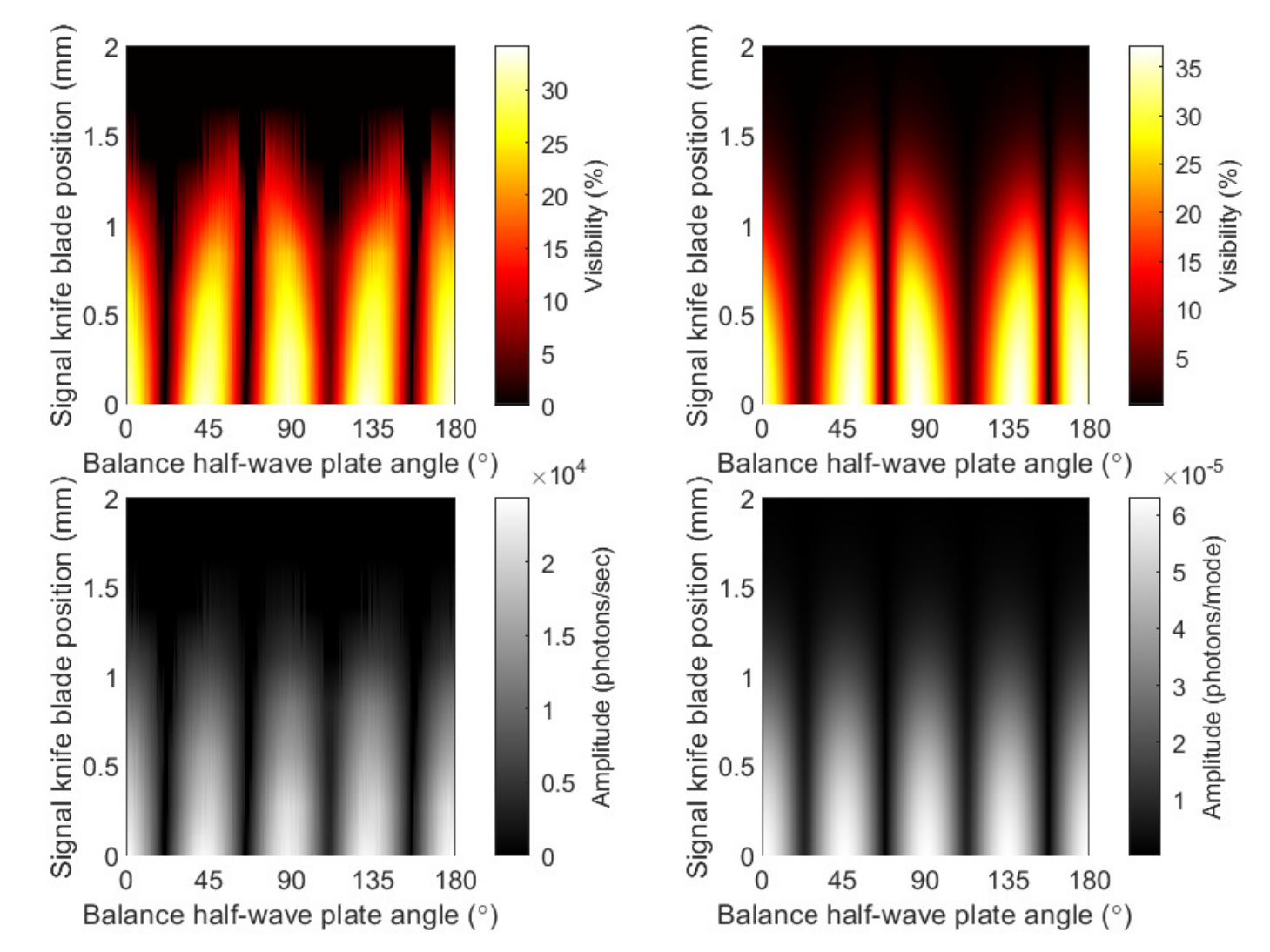}
\caption{Left: experimental data showing the visibility (top) and amplitude (bottom) of the interference fringes as a function of both power balance between the two crystals (adjusted via half-wave plate angle, x-axis), and primary signal loss (introduced via a stepper mounted knife blade, position marked by y-axis). Right: a theoretical model of experimental design.}
\label{fig:signalresults}
\end{figure}

Experimental results on how the visibility and amplitude of the interference fringes change as a function of power balance (half-wave plate angle) and first signal beam losses (introduced by the knife blade position) are shown in the left panels of Figure\,\ref{fig:signalresults}. Each data point of visibility was extracted from a fit to a sine wave for data taken such as that shown in the right panel of Figure 3, where the first idler phase has been scanned. Due to an overestimation of the visibility at very low count rates--- resulting from a highly varying noise floor---the visibility was clamped to zero in cases where the signal-to-noise ratio (SNR) $<1$; this gives rise to the sharp edges seen in the experimental data plots. SNR was calculated from the visibility goodness-of-fit: $R^2/(1-R^2)$). As is immediately evident, the high visibility cannot be maintained at high losses---even at extreme power unbalancing---due to the reduced SNR as a result of fewer signal photon counts.  In the experimental visibility plot (top left) there is a distinct asymmetry in the data as seen in the theory presented in the top right panel of Fig.\,\ref{fig:theory}. Indeed, using this theory and entering realistic numbers for fixed losses, beam diameters, beam-splitter extinction ratios, and background counts (0.01\,Hz), yields the theoretical results presented in the right panel of Fig.\,\ref{fig:signalresults}. The match between theory and experiment in shape and visibility is only reached with an additional 60$\%$ loss to the first idler beam.

\begin{figure}
\includegraphics[width=8.3cm]{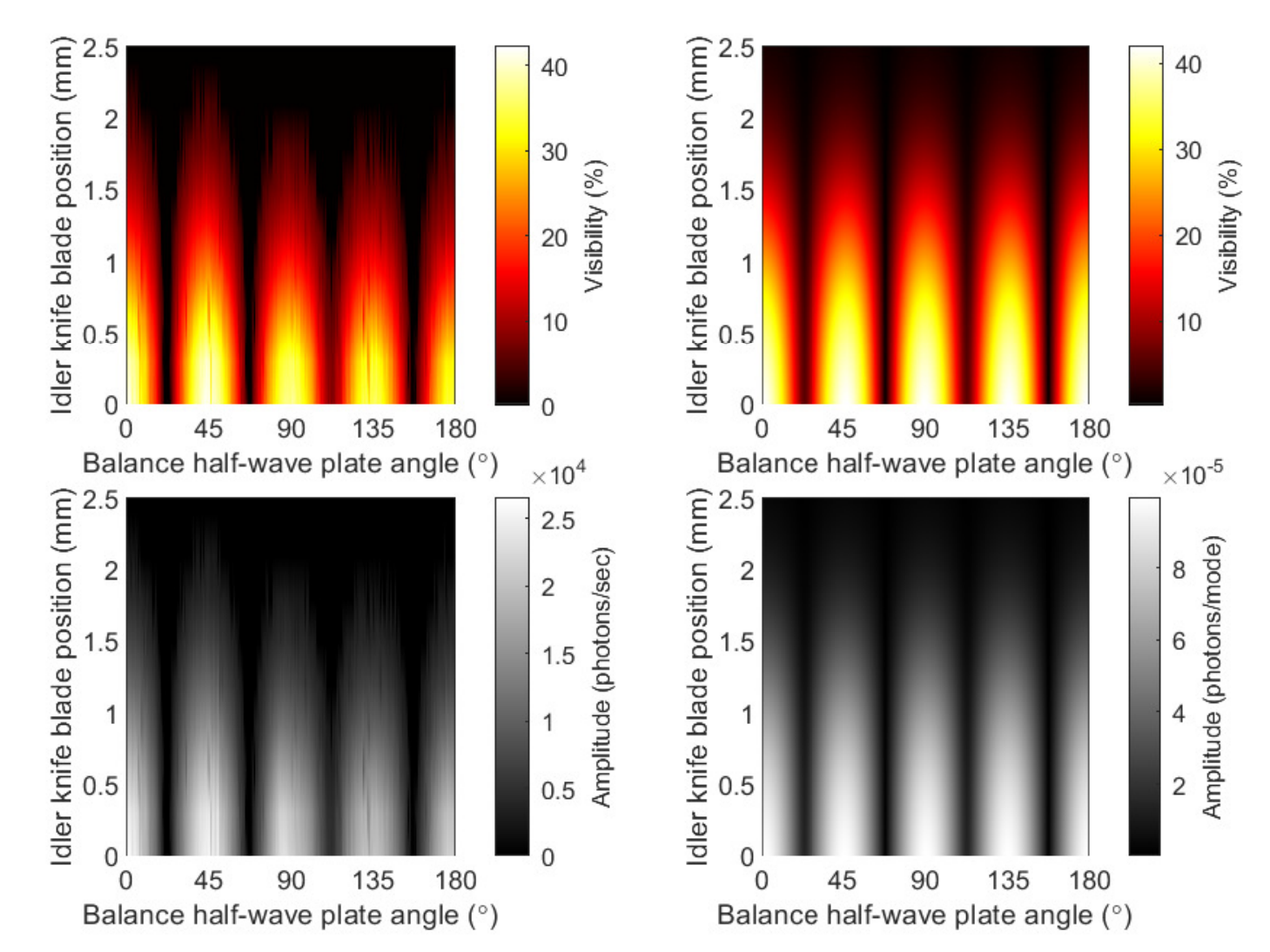}
\caption{Left: experimental data showing the visibility (top) and amplitude (bottom) of the interference fringes as a function of both power balance between the two crystals (adjusted via half-wave plate angle, x-axis), and primary idler loss (introduced via a stepper mounted knife blade, position marked by y-axis). Right: a theoretical model of experimental design.}
\label{fig:idlerresults}
\end{figure}

This match between theory and experiment confirms the validity of the model, though higher initial experimental fringe visibility would allow for regimes of even higher signal losses to be probed. 
Figure\,\ref{fig:idlerresults} shows the same plots as Fig.\,\ref{fig:signalresults}, but for varying internal idler losses (position of the knife blade in the idler beam). As predicted by the model, the data shows no sign of the asymmetry present when there is high first signal beam loss. Again, data fitted with an SNR $<1$ has been clamped to zero to mitigate over-fitting visibility values at low SNR. The symmetry shown in this plot confirms that the low maximum visibility observed in this system is not due to first signal beam losses, but instead may be due to high first idler beam loss. This loss (or other forms of distinguishability such as polarization rotation) is difficult to quantify but could be mostly due to misalignment and possible optical aberrations within the setup.

\begin{figure*}
\includegraphics[width=13cm]{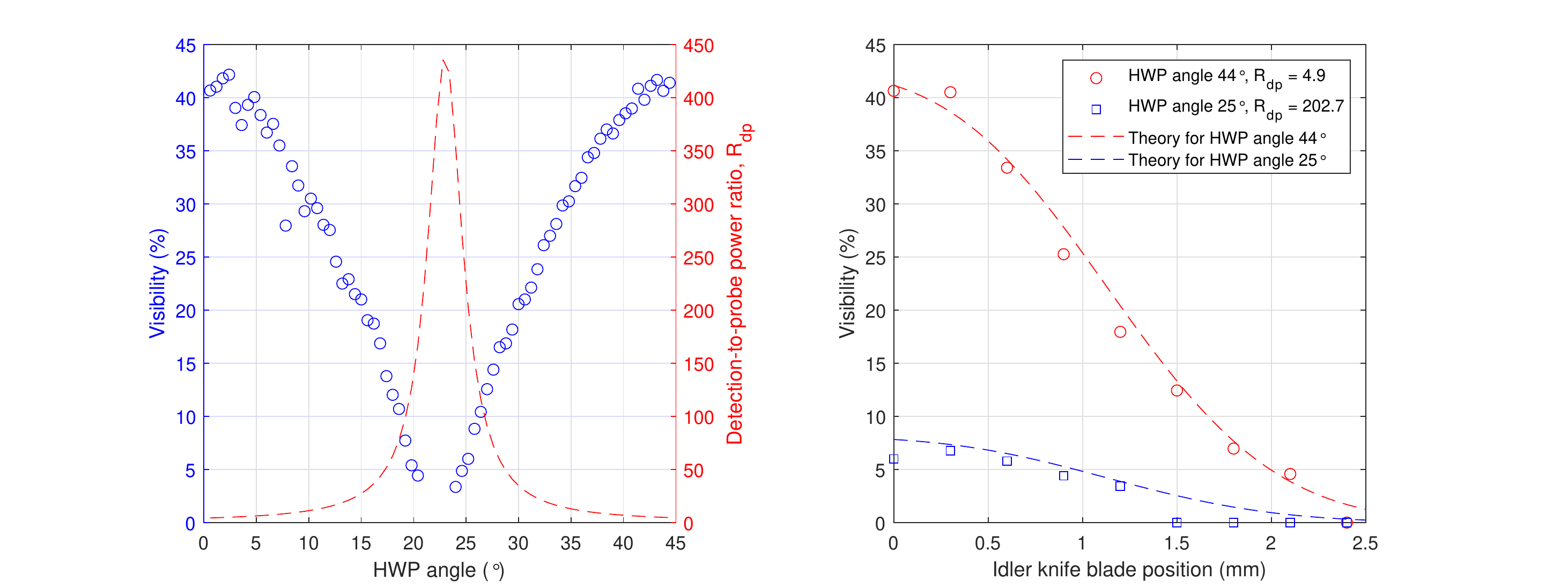}
\caption{Left: Visiblity (without knife blade losses) (left axis) and detection to probe power ratio, $R_\text{dp}$ plotted as a function of balance half-wave plate angle. Right: Visibility plotted as a function of losses in the primary idler arm for two different half-wave plate positions: 44$^{\circ}$ (red squares) and 24$^{\circ}$ (blue circles). Lines show the theoretical model. }
\label{fig:ifimage}
\end{figure*}

The data presented in Fig.\,\ref{fig:idlerresults} represents how typical `imaging with undetected photons' systems operate, where losses in the idler arm are mapped to changes in the visibility of the interference fringes of the signal photons. An obvious interpretation of these results is that perfect balancing gives the optimal performance for this kind of imaging system. However, such a conclusion would ignore the advantages of `interaction-free' operation discussed above. As stated in the theory section above, such a discrepancy in the probing and detection optical powers offers the opportunity to observe delicate processes in far greater detail than ever before. Figure\,\ref{fig:ifimage} demonstrates this principle: with the balancing half-wave plate set at 44$^{\circ}$, the visibility is at its peak and shows well the reduction with respect to losses (position of the knife blade). While the visibility has been severely reduced at a half-wave plate angle of 24$^{\circ}$, the ratio of detected-to-probing power has increased by a factor of $>40$ to over 200 (for a power balance ratio $R_\text{p}$ of 0.01). Such numbers mean that for every nW of power detected the sample can be exposed to $<5$\,pW of illumination power (compared to 205\,pW at 45$^{\circ}$). For the data presented here, the average signal count rate is $3\times10^{4}$ cps, corresponding to 3.8\,fW, giving the power incident on the sample (at a balance half-wave plate position of 24$^\circ$) as 18.7\,aW.

\section{\label{sec:level5}Conclusions}
We have shown that the balance of pump power in a nonlinear interferometer can be used to improve interference visibility in the presence of loss of the first signal beam. The result of this interplay of interferometer balancing and losses is key to understanding and implementing real-world imaging systems based on nonlinear interferometry. Furthermore, we have demonstrated that an imbalance can also be purposefully introduced at the expense of visibility to maximise the power unbalance in the probing and detecting beams. This allows a sample to be illuminated with fewer photons and significantly lower power than that detected, resulting in an ultra-low power imaging methodology. Such a situation allows the construction of imaging systems capable of probing samples effectively `in the dark' which could include photosensitive molecules, biological samples, or historical artefacts. This technique opens the door to the possibility of imaging in the mid-infrared far below the noise floor of current mid-IR camera technology through the transferal of information from a low power longer wavelength beam, to one of a higher power, and shorter wavelength. 

\begin{acknowledgements}
We acknowledge funding from the UK National Quantum Hub for Imaging (QUANTIC, No. EP/T00097X/1), an EPSRC DTP, and the Royal Society (No. UF160475).
\end{acknowledgements}




\bibliography{References}

\end{document}